\documentclass[a4paper,12pt]{article}
\usepackage{fullpage}
\usepackage{ifpdf}
\usepackage{cite}
 \usepackage[dvips]{graphicx}
\usepackage[cmex10]{amsmath}
\usepackage{amssymb}
\usepackage{algorithmic}
\usepackage{array}
\usepackage[tight,footnotesize]{subfigure}
\usepackage[caption=false,font=footnotesize]{subfig}
\usepackage{url}
\usepackage{tikz}
\usepackage{comment}
\usepackage{booktabs}
\usepackage{dcolumn}
\usepackage{setspace}
\newenvironment{proof sketch}[1]{\noindent {\emph{Proof sketch of #1:}}}{\hfill \qed}

\newcommand{\bw}{\textrm{bw}}

\newcommand{\inn}{\mathsf{in}}

\newcommand{\din}{d_{\inn}}

\newcommand{\etal}{{\it et~al.}}

\newcommand{\RR}{\mathbb{R}}

\begin{document}

\begin{titlepage}
  \title{Algorithms for Network-on-Chip Design with Guaranteed QoS} \author{ Guy Even\thanks{ School
      of Electrical Engineering, Tel-Aviv University, Email: \protect\url{guy@eng.tau.ac.il}
}
    \and Yaniv Fais }

 \maketitle
  \begin{abstract}
    We present algorithms that design NoCs with guaranteed quality of service.  Given
    a topology, a mapping of tasks to processing elements, and traffic requirements
    between the tasks, the algorithm computes the interconnection widths, a detailed
    static routing, and a periodic scheduling.  No headers, control messages, or
    acknowledgments are required.

    The algorithm employs fractional Multi-Commodity Flow (MCF) that determines the
    widths of the interconnections as well as the routes of the flits.  The MCF is
    rounded to a periodic TDM schedule which is translated to local periodic control
    of switches and network interfaces.

    Our algorithm is applicable to large instances since every stage is efficient.
    The algorithm supports arbitrary topologies and traffic patterns.  Routing along
    multiple paths is allowed in order to increase utilization and decrease latency.

    We implemented the algorithm and tested it with the MCSL benchmark. Experiments
    demonstrate that our solution is stable and satisfies all the real-time
    constraints.
  \end{abstract}
\end{titlepage}

\section{Introduction}
Networks-on-Chips (NoCs) have been proposed as a way to design Multi-Processor
Systems-on-Chip (MPSoC)~\cite{dally2001route,seitz1990let}. NoCs constitute good
engineering practice in the sense that they support modularity and, even more
importantly, NoCs help decouple communication from computation and processing.  NoCs
are characterized by stringent real-time constraints and high reliability (packets can not
be lost). Satisfying these requirements forces one to resort to short packets (called
flits) and, even more importantly, use routing and scheduling without the layers of
abstraction that characterize networks (such as TCP/IP).

A common paradigm in network design in general, and NoCs in particular, is the usage
of virtual channels and dynamic routing~\cite{goossens2010aethereal}.  Virtual
channels provide an abstraction of end-to-end links which allow for time-division
multiplexing (TDM) of the network interconnections. Dynamic routing offers
flexibility in using the network resources.  Successful online allocation of virtual
channels cannot be guaranteed~\cite{awerbuch1993throughput,aspnes1997line}.  Hence,
online routing schemes cannot assure reliability let alone Quality-of-Service (QoS).
As partial provision of bandwidth in real-time systems renders the system useless,
the design of NoCs for a real-time applications requires prior knowledge of the traffic.

In this paper we focus on designing NoCs with respect to persistent traffic.
Specifically, we address the following issues that are usually not addressed in
theoretical papers: avoiding headers to reduce overhead, guaranteeing arrival of
packets so that control messages (such as acknowledgments) are not needed, and
guaranteeing real-time constraints.

\subsection{Contribution}
We present an efficient methodology for designing NoCs.  The methodology can be
applied to any NoC topology and deals in a unified fashion with the following
algorithmic problems: assigning widths to interconnections in the NoC, computing
routing, computing the scheduling, and computing the local controls of each network
component.  Our methodology can be used to automatically generate NoCs including
detailed design of all the network components. Such a tool can help in 
implementing  specific real-time applications over parallel platforms such as
multi-processor systems on a chip.

The starting point is a traffic pattern that ``dominates'' the actual traffic to be
supported.  The first stage of our algorithm solves a fractional multicommodity flow
(MCF) problem.  The MCF formulation is used to compute the required edge capacities
and is easily adapted to different objective functions such as: reducing total wire
cost, bounding the capacity of links, and limiting lengths of flow paths.  The
computed MCF determines edges capacities as well as the routing of the requests.  A
periodic schedule that supports the MCF is computed in the next stage.  This periodic
schedule is then implemented by programming the control of the nodes in the network.

In NoCs, the common practice is to send (long) messages by long trains of contiguous
flits~\cite{goossens2010aethereal}.  We deviate from this practice because such long
trains block the network, incur large delays, and reduce  utilization. Instead, we
employ schedules in which flits belonging to the same message can be sent in
non-contiguous sets of time slots and even along different paths.

We show that periodic schedules can be implemented without any headers, control
messages, nor acknowledges. Our synchronous implementation delivers all the packets
(no packet is dropped) and does not require ``warming up'' of the network. We
guarantees safe delivery of every packet. Our implementation suggests a trade-off in
which local control (an abundant resource in modern VLSI chips) can help save in wires
(a nonscalable resource).

To increase network utilization we remove the restriction that all packets of the same
session must  traverse the same path.  Splitting flow into multiple paths helps
balance the load, but is considered unfavorable due to packets arriving
out-of-order. Our implementation of periodic schedules reorders the packets without
requiring the attachment of serial numbers to packets.

We also suggest a method for designing NoCs for real time systems. The real-time
system is modeled by a constraints task communication graph (TCG) that specifies
dependencies between tasks. In addition, the time required to complete
each task (once its inputs are available) and the length of the messages between
tasks is also part of the TCG. We suggest to bound the delay of each message by a
linear function of the message length. Thus, the choice of the linear function of the delay
bound implies deadlines for all tasks, and we obtain a real-time benchmark that can be
easily checked.

We show how to reduce the delay bounds of the messages in the TCG to a
multi-commodity flow problem. This reduction enables us to employ our NoC design
methodology to support real-time systems. Our simulations suggest that the
methodology yields NoC designs that meet the deadlines, and thus real-time
performance is achieved.

\subsection{Previous Work}

Multi-commodity flows were previously used to design NoCs. In
~\cite{murali2004bandwidth,bertozzi2005noc} multi-commodity flow was used for
computing mappings of cores in NoC architectures.  Schoeberl \etal\
\cite{schoeberl2012statically} employed an integer multi-commodity flow for designing
NoCs with uniform traffic. Murali \etal~\cite{murali2006multi} solve a fractional MCF
and  randomly round the fractional flow to obtain an unsplittable flow solution. Hu
\etal~\cite{hu2005physical} employ MCF to reduce the power consumption of NoCs.
Hu \etal~\cite{hu2006communication} employ MCF to find the best topology from a
given set for a latency-power product objective.

S{\o}rensen \etal\ \cite{sorensen2014metaheuristic} present a meta-heuristic scheduler for
inter-processor communication in multi-core platforms using time division multiplexed
(TDM) NoCs. Given a specification of the target platform and of the application, the
algorithm generates a periodic TDM schedule. The algorithm is based on solving an
integer multicommodity flow problem. Since this problem is NP-Complete, the running
time of the algorithm does not scale well. In addition to designing a more general
scheduler, the goal of minimizing the period of the schedule is obtained by
increasing the frequency of the TDM clock. The results obtained
in~\cite{sorensen2014metaheuristic} for the MCSL benchmark on the $8\times 8$ torus
indicate\footnote{Our guess is that the results were reported for a mesh with links
  $32$-bits wide.}  that limiting the TDM period to $100$ time slots requires
increasing the TDM clock frequency by a factor of $10$ to $100$.

Andrews et. al~\cite{andrews2000general} prove bounds on packet delays for
periodic scheduling in store-and-forward packet routing networks. (We elaborate on
their setting in Sec.~\ref{sec:token}.) They proved the existence of a periodic
schedule with an asymptotic optimal delay bound. They also present a simple
randomized algorithm in which the asymptotic delay bound is bigger by a logarithmic
factor.  Their setting assumes that paths and rates for all sessions are given as well as
a positive constant slack in the congestion of the edges.  The result of Andrews
et.~\cite{andrews2000general} justifies focusing on periodic schedules in this
setting.

\subsection{Organization}
In Sec.~\ref{sec:what}, we briefly describe what NoCs are.
In Sec.~\ref{sec:token}, we discuss the token based periodic schedules of Andrews
et. al~\cite{andrews2000general}.
In Sec.~\ref{sec:spec}, we propose a parametrized timing specification of real-time
systems.
In Sec.~\ref{sec:implementation}, we show how to implement periodic schedules without
requiring any headers.
In Sec.~\ref{sec:method}, we present our NoC design methodology.
In Sec.~\ref{sec:MCF}, we present a reduction of real-time task communication
graphs to multi-commodity demands.
Experimental results are described in Sec.~\ref{sec:exp}. Further directions of
research are discussed in Sec.~\ref{sec:further}.
\section{What is a Network-on-Chip?}\label{sec:what}
Networks-on-Chips (NoCs) are store-and-forward packet networks that support
communication between modules on a chip (e.g., multiple cores). The idea is to
replace dedicated point-to-point interconnections between the modules as well as
buses by a packet forwarding network.  As in regular data networks, the NoC hides the
details of how data is sent between modules.

A NoC can be modeled by a graph $G_N=(P\cup V, E)$, where $P$ denotes the set of
processing elements (PEs) and $V$ denotes the set of NoC switches. Each PE is
connected to the NoC via a dedicated network interface (NI), hence we identify each
PE with its network interface.

The design of a NoC involves classical issues in parallel computation and networking,
including: choosing the network topology, mapping tasks to PEs, computing the routes
of packets, scheduling packets, managing buffers, etc.

\section{Periodic Schedules Based on Token Sequences}\label{sec:token}
In this section we overview the periodic schedules that are computed for
store-and-forward packet routing networks in Andrews et.
al~\cite{andrews2000general}.  We are given a synchronous network $N$ with arbitrary
topology. The goal is to route sessions over $N$.  Each session $i$ consists of a
source node $a_i$, a destination node $b_i$, a simple path $p_i$ from $a_i$ to $b_i$,
and an injection rate $r_i$. The injection rate limits the number of packets that may
be injected by session $i$. Formally, for every $t$, at most $r_i\cdot t +1$ packets
may be injected by session $i$ during every interval of length $t$. The sessions must
satisfy the following congestion property.  There exists an $\varepsilon>0$ such that
for every edge $e$, $\sum_{p_i\ni e} r_i \leq 1-\varepsilon$.

In the language of multicommodity flows, each session is a commodity with demand
$r_i$. All edges have unit capacity, flow is not splittable, and the routing is given.
The algorithm requires that the load of every edge may not exceed $1-\varepsilon$.

A template-based periodic schedule is suggested as a basis for the scheduling
algorithm in~\cite{andrews2000general}.  The terminology of this schedule consists of
a \emph{period}, \emph{tokens}, \emph{token sequences}, and \emph{templates}.  Let
$\Phi$ denote the \emph{period} of the schedule.  Each directed edge in the network
is attached a \emph{template} that is a table with $\Phi$ entries.  Each entry in a
template is called a \emph{token} and its value is a serial number of a token
sequence.  A \emph{token sequence} for session $i$ is an allocation of one token in
each template of the edges along the path $p_i$. The token sequence is used for
sending packets at a rate of one packet per period from the source $a_i$ to the
destination $b_i$.  To support the demand of session $i$, session $i$ should have
$r_i\cdot \Phi$ token sequences assigned to it.

The rule for scheduling packets according to a template is as follows. Consider the
packets of session $i$. The packets wait in a FIFO input queue of session $i$ in the
source node $a_i$.  Let $e_1$ denote the first edge of $p_i$.  Whenever a time slot
occurs in which the token in the template of $e_1$ is a token sequence of session $i$,
the first packet in the input queue of session $i$ is dequeued, and sent along the
edge $e_1$. From this point on, each further hop of the packet may only use the same
token sequence used for the first hop. This means that the incoming packet is stored
in the interior node together with its token sequence. This packet is sent along the
next edge of $p_i$ as soon as a time slot arrives whose token equals the token
sequence of the packet.

A naive implementation of the scheduling algorithm requires that each packet carry
the token sequence that was used for its first hop (in Sec.~\ref{sec:implementation}
we show that this is not required). Alternatively, ~\cite{andrews2000general} present
a distributed randomized scheduling algorithm. The distributed algorithm delays each
token independently and uniformly. Each packet is preassigned a deadline for each
hop.  Contention is resolved by employing an earliest-deadline-first policy (EDF).
The distributed algorithm does not require templates.  However, the header of each
packet must contain an encoding of its path (such as its session number so that
each intermediate node can determine the next hop) as well as its random delay (so
that EDF can applied).

\section{Timing Specification of Real-Time Systems}\label{sec:spec}
In this section we propose a parameterized timing specification of a real-time
system. The specification is based on a precedence graph over the tasks and bounds
the delay of messages by a linear function. This specification is used as a benchmark
in the evaluation of real-time applications implemented over NoCs.

\paragraph{Task Communication Graphs (TCGs).} Real-time systems can be described by a
precedence graph over tasks.  Dependencies between tasks are derived from the
communication between the tasks.  The precedence graph is a directed acyclic graph,
called a \emph{task communication graph}. Each vertex in the TCG corresponds to
a task and is labeled by the time it takes the task to complete once all the inputs
are ready.  An arc from task $u_i$ to task $u_j$ means that task $u_i$ sends a
message to task $u_j$.  A task $u_j$ cannot begin before all the messages from
incoming edges arrive.  Each arc in the TCG is labeled by the length of the message.
External inputs are modeled by sources in the TCG and external outputs are modeled by
sinks. Each source is labeled by an \emph{arrival time} that specifies when this
input is injected into the system. Each sink is labeled by a \emph{deadline} that
specifies the time by which it should receive its incoming message (which is the
output of the system).  Real-time compression of video from a camera is a canonical
example of a real-time system. Arrival times are determined by the rate in
which raw data from the camera is fed, and deadlines of compressed frames are
determined by the maximum allowed delay. 

\paragraph{Communication delay bounds: from edges to tasks.}  We propose to model the
delay of a message by a linear function. Let $|m|$ denote the length of a message,
$L$ the latency parameter, and $\alpha$ denote the width of the edge $e$.  The delay
bound of a message $m$ sent along $e$ is at most $L+|m|/\alpha$ time slots.  Note
that higher values of $L$ or lower values of $\alpha$ allow more time for
communication, and thus postpone deadline.  The completion times of each task are
computed by dynamic programming based on the task communication graph, message
lengths, computation duration of each task, input arrival times, and the
communication delay parameters $L$ and $\alpha$.

\paragraph{Satisfying real-time specification.}  A pair $L$ and $\alpha$ is 
\emph{feasible} if the completion time of the sinks does not exceed their
deadlines. In the absence of deadlines in sinks (as well as intermediate nodes), the
completion times of such a hypothetical network serve as our benchmark.  Namely, we
define the deadlines to be the completion times of the tasks if each message $m$ is
delivered after $L+|m|/\alpha$ time slots. The role of the parameters $L$ and
$\alpha$ is to obtain deadlines if we have none, and to help in the reduction of the
TCG specification to a multi-commodity flow problem (see Sec.~\ref{sec:MCF}).  Our goal
is to design a NoC that delivers the same performance as such a hypothetical network.
As $\alpha$ increases and $L$ decreases, the challenge in designing a NoC is harder
because the ``competing'' hypothetical network has a smaller communication delay.

A system meets the real-time specification if the outputs satisfy the deadline
constraints (that depend on $L$ and $\alpha$).  The \emph{lag} of a task is the
difference between the time in which the task ends and the time in which it is
supposed to end.  Positive lags correspond to violations of deadlines. In fact,
bounded lags (that do not grow as a function of time) are usually tolerable.
Unbounded lags that increase linearly over time, often called \emph{drifts}, and mean
that the system is running too slowly. Elimination of drifts (using the same network)
requires accelerating the clock rate of the network.

\section{Implementation of Periodic Schedules}\label{sec:implementation}
In this section we show how to implement the switches and network interfaces in a NoC
so that they execute a given periodic schedule. This implementation does not require
any headers. In particular, a packet does not need to carry its path, session number,
serial number, etc.

\subsection{Representation}
We first describe how a periodic schedule is represented. One option is to have a
schedule based on token sequences. Another option is that the periodic schedule is
represented by a set of periodic schedules, one per edge. Let $S$ denote the set of
sessions  and let $\sigma_e$ denote the schedule for edge $e$. The
schedule of an edge $e$ is a function $\sigma_e: \{0,\ldots, \Phi-1\} \rightarrow S$
that specifies which session is scheduled in each time slot.

\paragraph{Reduction to token sequences.}
We reduce a periodic schedule that is represented by edge schedules to token
sequences by matching incoming slots and outgoing slots of each session. By following
time slots that are matched along the flow paths, we can ``decompose'' the flow of
packets of each session into token sequences. We note that such matchings are
possible because flow must be conserved in each intermediate node.  Namely, the
number of time slots allocated to $s$ along incoming edges equals (or is at most) the
number of time slots allocated to $s$ along outgoing edges.  Hence such a matching
exists. In this matching it is desirable to minimize the delay from the source to
destination of session $s$.  If session $s$ uses a single path, then the matching is
based on a simple first-in-first-out policy. We point out that computing matchings
that minimize end-to-end delay for splittable flows is an interesting problem.

\subsection{Implementation}
The proposed NoC architecture (which is similar
to~\cite{murali2004bandwidth,sorensen2014metaheuristic}) consists of
processing elements (PEs), network interfaces, switches (with small buffers and local
control) and interconnections (whose widths are determined by the algorithm). Each PE
is connected to a switch via a Network Interface (NI) (for simplicity, we allow each
NI to be connected only to a single switch).

Once we have a representation by token sequences, each switch node $v$ is implemented
by a memory and a control as follows. The memory can store $\Phi\times \din$ flits,
where $\din$ is the number of incoming edges.\footnote{In our implementation, we
  reused memory if a flit stays in the memory for only one time slot.} In time slot $i$, the memory stores all
the incoming flits in the $i$th row of the memory. Each outgoing edge $e$ is
controlled by a function $f_e : \{0,\ldots,\Phi-1\} \rightarrow \{0,\ldots,
\Phi-1\}\times \{1\ldots,\din\}$. The value of $f_e(t)$ specifies the memory address 
that contains the flit to be sent along edge $e$ in time slot $t$. Note that a
concurrent read and write to the same location mean that the read returns the value
before the write (because a flit cannot arrive and be forwarded in the same time
slot).

Network interfaces (NIs) are sources and sinks of sessions (but not intermediate
nodes).  An NI contains a queue for each session. If the NI is the source of the
session, then it dequeues a packets whenever the time slot of the outgoing edge is
allocated to the session, and sends the packet along this edge. If the NI is the destination
of the session, then it adds the arriving packet to the queue.

\subsection{Reordering of Packets}
If a session is routed along multiple paths, then the arrival order of the packets
may not equal the order in which they have been sent. We describe now how the network
interface of the destination can reorder the packets without having to attach serial
numbers to the packets.

As described above, the NoC implements the policy dictated by a periodic schedule
that is represented by token sequences.  Consider session $s$ and assume that the
schedule has $k$ token sequences for session $s$. Suppose we number the token
sequences from $1$ to $k$, and let $t_i$ denote the time slot in which token sequence
$i$ leaves the source. We assume that $t_1\leq t_2\leq \cdots \leq t_k$.  (If
multiple token sequences leave the source in the same time slot, then we order them
arbitrarily, but the source must send the packets according to this order.)  The
destination network interface of session $i$ reorders the incoming packets as
follows.  Note that packets that are delivered using the same token sequence (but in
different periods) arrive in order.  Hence, the sorting needs only to deal with
merging $k$ sorted sets.  Let $D_i$ denote the accumulated delay of token sequence
$i$. This means that a packet starts moving in time slot $\Phi\cdot \ell +t_i$ must
arrive at time slot $\Phi\cdot \ell +t_i +D_i$.  Hence the difference in time slots
between the arrival of consecutive packets from token sequences $i$ and $i+1$ equals
$t_{i+1}+D_{i+1} - (t_i+D_i)$. As this difference is fixed, the network interface
can easily merge the packets from different token sorted in the correct order.

\section{NoC Design Methodology}\label{sec:method}
In this section we present a methodology for designing NoCs. The starting point is
the NoC topology and multi-commodity demands over the processing elements of the NoC.
The algorithm computes the widths of the interconnections in the NoC and  the
periodic schedules of NoC component (switches and network interfaces). Based on these
widths and schedules, the network can be automatically generated. 
The algorithm proceeds in the following phases.

\paragraph{Fractional multicommodity flow (MCF) formulation.}  The MCF instance is over
the NoC graph $G_N=(P\cup V, E)$ with demands $\bw(PE_i,PE_i)$ between pairs of PEs.
The demands describe the traffic rates between the PEs.  The algorithm computes a
flow $f_{(PE_i,PE_j)}$ over the NoC graph for each ordered pair $(PE_i,PE_j)$.  The
flow amount of $f_{(PE_i,PE_j)}$ equals the demand $\bw(PE_i,PE_j)$.  Let
$f(e)\triangleq \sum_{i,j} f_{(PE_i,PE_j)}(e)$ denote the total flow along $e$.  The
width of the interconnection $e$ is proportional to $f(e)$.  Linear programming
formulations of the MCF are quite flexible, so we describe three possible objectives.
\begin{enumerate}
\item Minimize total interconnection cost. In this formulation each edge $e$ has a
  cost $c(e)$ that corresponds to its length. The goal is to minimize $\sum_e
  c(e)\cdot f(e)$.  In fact, this instance uses shortest paths routing and can be
  solved without linear programming. The solution is simply an overlay of shortest
  paths of all the demands.
\item Minimize maximum congestion along edges. In this formulation each edge $e$ has
  a width $u(e)$. We add constraints of the form $f(e)\leq \lambda \cdot u(e)$ for
  each edge. The objective is to minimize $\lambda$.
\item Limit the lengths of flow paths. The length of a flow path is a heuristic
  indication of the cost and delay associated with the path. 
\end{enumerate}
We remark that linear programming solvers can easily solve instances with thousands
of commodities.

\paragraph{Rounding of flows.}
Let $\Phi$ denote the requested period of schedule.\footnote{There is a trade-off in
  the choice of $\Phi$. The larger $\Phi$ is, the finer granularity we obtain, and
  thus rounding incurs a smaller overhead. On the other hand, the cost of the NoC
  switches increases as $\Phi$ grows. }  If flow is measured in flits per time slots,
then the smallest unit of flow (i.e., granularity) is $1/\Phi$.  This means that we
need to round the flows of each pair to multiples of $1/\Phi$ as follows. For each
pair $(PE_i,PE_j)$, decompose the flow $f_{(PE_i,PE_j)}$ into flow paths
$p_1,p_2,\ldots$.  Assume that the paths are sorted in descending flow amount,
namely, $f_{(PE_i,PE_j)}(p_k) \geq f_{(PE_i,PE_j)}(p_{k+1})$.  Initialize $d(i,j)$ to
equal the demand $\bw(PE_i,PE_j)$.  Scan the paths $\{p_k\}_k$ starting with $p_1$.
If $d(i,j)>0$, round up $f_{(PE_i,PE_j)}(p_k)$ to a multiple of $1/\Phi$ and update
$d(i,j)\gets d(i,j)-f_{(PE_i,PE_j)}(p_k)$.  If $d(i,j)=0$, then the remaining flow
paths are erased.

\paragraph{Determine interconnection widths.}
If flow is measured in flits per time slots, then the width $w(e)$ (in bits) of each
interconnection $e$ in the NoC graph is simply $\lceil f(e) \rceil \cdot k$, where
$k$ denotes the flit size. Thus the width of each interconnection is big enough to
accommodate the planned flow along it and there is no need to split flits.

\paragraph{Periodic Scheduling.}  A greedy allocation of time
slots to demand along the edges is guaranteed to work. One can apply the algorithm of
Andrews et. al~\cite{andrews2000general} to compute a periodic schedule with proven
bounds on the delay of each packet.

\section{Reduction From TCG to Multi-Commodity Flow}\label{sec:MCF}
In this section we present a reduction of a specification of a TCG of a real-time
system to a multi-commodity flow problem.  This reduction enables us to employ our
NoC design methodology to support real-time systems.

Consider a TCG $G_T = (T,E_T)$ over a set of tasks $T$. The edges are labeled by the
length of the messages. The delay bound of a message $m_e$ that corresponds to $e$ is
$L+|m_e|/\alpha$, where $L$ denotes the latency parameter and $\alpha$ denotes the width
of the virtual channel.

Our goal is to define a multi-commodity flow (MCF) problem for a NoC that is supposed
to support the execution of the real-time system modeled by the TCG $G_T$.  The NoC
is a graph $G_N=(P\cup V, E)$, where $P$ denotes the set of processing elements
(PEs) and $V$ denotes the set of NoC switches (we identify each PE with its network
interface,  so we do not need to model network interfaces here).  We are given a
mapping $\pi:T\rightarrow P$ that assigns each task to a PE. 

The MCF demands are set of demands between pairs of PEs. Namely, $\bw: P\times P
\rightarrow \RR$, where $\bw(PE_i,PE_j)$ denotes the maximum rate of traffic from the
processing element $PE_i$ to $PE_j$.  Fix $PE_i$ and $PE_j$. We suggest to define the value of
$\bw(PE_i,PE_j)$ as follows.  Consider all the edges $e=(\tau_a,\tau_b)$ in the TCG
such that $\pi(\tau_a)=PE_i$ and $\pi(\tau_b)=PE_j$. Each such edge $e$ induces a
flow demand $f_e$ from $PE_i$ to $PE_j$ during an interval $I_e$. The interval $I_e$
starts when $\tau_a$ is completed, and its length is $L+|m_e|/\alpha$, where
$|m_e|$ is the length of the message sent along $e$.  The flow amount $f_e$ equals
$|m_e|/\alpha$.  For every time $t$, let $f(t)$ equal the sum of the induced flow demands
$f_e$ for which $t\in I_e$. We define $\bw(PE_i,PE_j)$ to be $\max_t f(t)$.

After defining the MCF demands, we may define the objective function of the MCF
according to Sec.~\ref{sec:MCF}, and apply the NoC design methodology.

\section{Experimental Results}\label{sec:exp}
\subsection{Benchmarks}
\paragraph{Random steady traffic.}  In the random traffic model we randomly choose pairs
of demands. The flit size is $4$ bits and the demand $\bw(P_i,P_j)$ (in flits) is chosen
uniformly from $\{0,\ldots,5\}$. Hence, on average, a virtual channel delivers $10$ bits per
period.  

The main purpose of the random steady benchmark is to test whether: (1)~the algorithm
is efficient for large instances, (2)~the algorithm computes routings with a
reasonable number of wires, and (3)~the algorithm computes schedules with high
utilizations, low latency, and small switch buffers.

\paragraph{The MCSL Benchmark~\cite{liu2011noc}.} The recorded traffic in the MCSL
benchmark is based on mapping a multi-threaded program to multiple cores (PEs). The
benchmark contains TCGs of applications such as decoding and encoding of Reed-Solomon
codes, FFT, etc. We used TCGs that were obtained from embeddings in $8\times 8$
meshes. We emphasize that the TCGs of the MCSL benchmark do not include deadlines of
external outputs or arrival times of external inputs.  This benchmark enables one to
perform detailed comparison between the real-time specification (deadlines of tasks)
and the completion times of tasks in the NoC-based system.

\subsection{Algorithm Implementation}
The algorithm uses the CLP COIN-OR LP solver~\url{https://projects.coin-or.org/Clp}. Even though the MCF
objective in our experiment is the sum of edge costs, we used the LP solver to verify
scalability.  The algorithm is implemented in C++.

\subsection{Simulator}
Our simulator is based on the HNoCS~\cite{ben2012hnocs} which uses the OMNET++
network simulator. This simulation is cycle accurate. We implemented generators for
several parameterized topologies ($3$-level Clos, Bene\v{s}, Fat Tree - Butterfly
with duplication of nodes, K-ary
N-fly~\cite{theocharides2005networks,bertozzi2005noc}, K-ary-N-Fly-Tree, and a random
graph).  We implemented modules for the network interfaces and the switches. We also
implemented modules that feed the traffic from the benchmarks to the network
interfaces. Simulation results are logged by monitoring modules.

\subsection{Results}

\paragraph{Random steady traffic.}  We executed a random traffic model with steady
traffic for $16$ to $256$ PEs over various topologies\footnote{The random NoC graph
  was obtained by a union of two perfect random matchings altered to obtain
  connectivity.}.  The schedule period is $8$ time slots.  The results of the
simulation are listed in Table~\ref{tbl:random}.  Basic floor-planning (for the
topologies that are not the mesh) was employed to estimate the length of each
interconnection as follows.  The PEs are organized in a square matrix, and each
switch is located in the ``center of mass'' induced by its distance to the PEs it
communicates with. The Euclidean distance between the locations of the endpoints of
an interconnection is used as an estimate of the interconnection length.  This
estimate of the length of interconnections is used as their cost for the MCF.  The
computed width of the interconnection is measured in bits (interconnections must be a
multiple of the flit size). The cost of an interconnection is the product of its
length and width. The buffers in the switches are referred to as the memory and their
size is counted in bits.  The latency of a flit between NIs is given in time slots
(i.e., clock cycles of the NoC).  The average utilization is the fraction of time
slots in which a message is delivered during the simulation (in the steady state).

These results demonstrate the flexibility of the algorithm with respect to various
topologies and many PEs. The running time of the algorithm for $64$ cores is about
two minutes. The average latency increases moderately as the number of cores
increases. For example, the average latency in the $4\times 4$ mesh is $3.11$ time
slots compared to a latency of $8.6$ in the $12\times 12$ mesh. Namely, the average
latency grows at a lower rate than the diameter.  In small networks, the mesh wins in
all parameters, however, in larger networks, the mesh has a higher latency (although
it remains cheaper as there are fewer switches).

\begin{table}
{\centering\includegraphics[width=\textwidth]{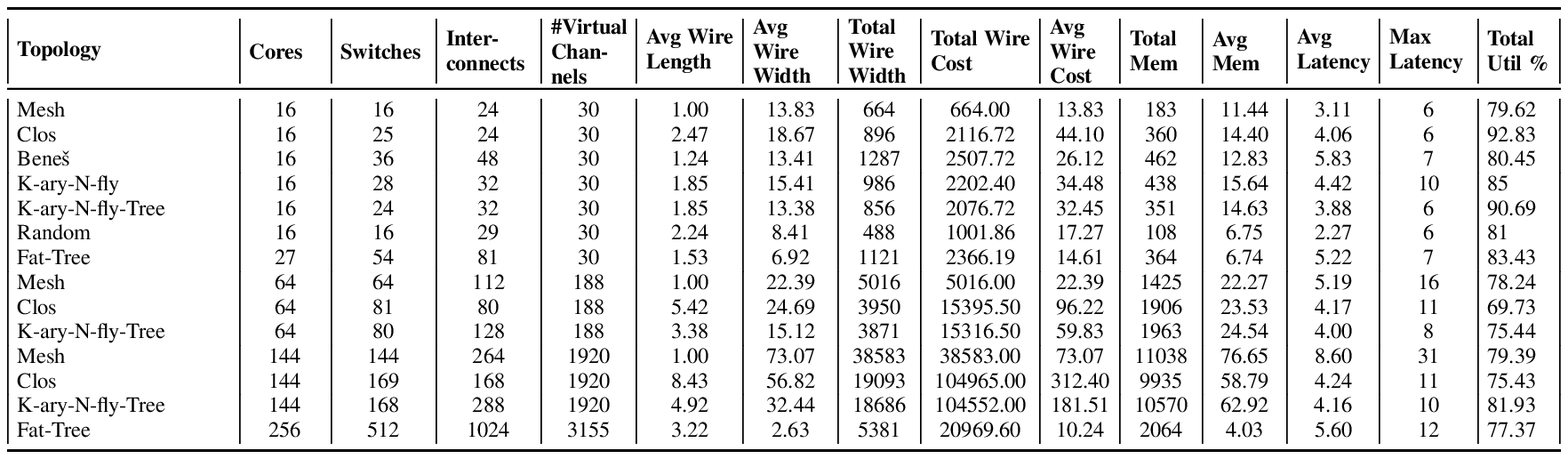}}
  \caption{Results with steady random traffic}
  \label{tbl:random}
\end{table}

\paragraph{MCSL Recorded Traffic.}  We executed the algorithm for the $8$ MCSL recorded
traffic patterns over an $8\times 8$ mesh. The period of the TDM schedule is $8$ time
slots. The real-time specification was derived from the TCG combined with an
end-to-end delay bound of $L=0$ (except for the FFT benchmark for which $L=8$ was
used) and a range of widths $\alpha\in\{4,8,16,\infty\}$ bits per second
($\alpha=\infty$ means that the real-time deadlines assume that any number of flits
can be sent simultaneously over every interconnection). The results of the
simulations are listed in Table~\ref{tbl:mesh 8x8}.  The results demonstrate that,
except for the FFT benchmark, the NoC supports the traffic patterns with reasonable
wire widths, buffer sizes, and latencies.  We elaborate below on the very low
utilization of time slots which is caused by the bursty nature of the traffic
patterns.

It is interesting to note the following remarks on the simulation results: (1)~ The
rounding of the fractional MCF to time slots and flits incurs an overhead of
$5\%\text{-} 10\%$.  This means that the rounding employed by the algorithm is rather
efficient, and that the cost of the solution with respect to the maximum bandwidth
requirements is at most $10\%$ higher than the optimal cost (for the benchmark
patterns).
(2)~Increasing the end-to-end delay $L$ of virtual links used to derive the real-time
specification has a small effect on the results of the algorithm. Only in the FFT benchmark
application did we use an end-to-end delay of $L=8$; in all the rest of the benchmark
applications we used $L=0$.
(3)~The real-time specification (based on the task communication graph, the running
time of each task, the lengths of the messages, and the parameters $L$ and $\alpha$)
defines the required end times of every task.  The simulation (based on the execution
of the tasks, the NoCs, and the periodic schedule) provides the end time of every
task. Thus, we can compute the lag of every task.  The sum of the lags over all the
tasks (as well as the lag of the last task) is negative for all the simulations
listed in Table~\ref{tbl:mesh 8x8}. In addition, the maximum lag is small. Hence, we
conclude that the simulated performance satisfies the real-time specification
throughout the execution of the applications in the benchmarks.

\begin{table}
{\centering\includegraphics[width=\textwidth]{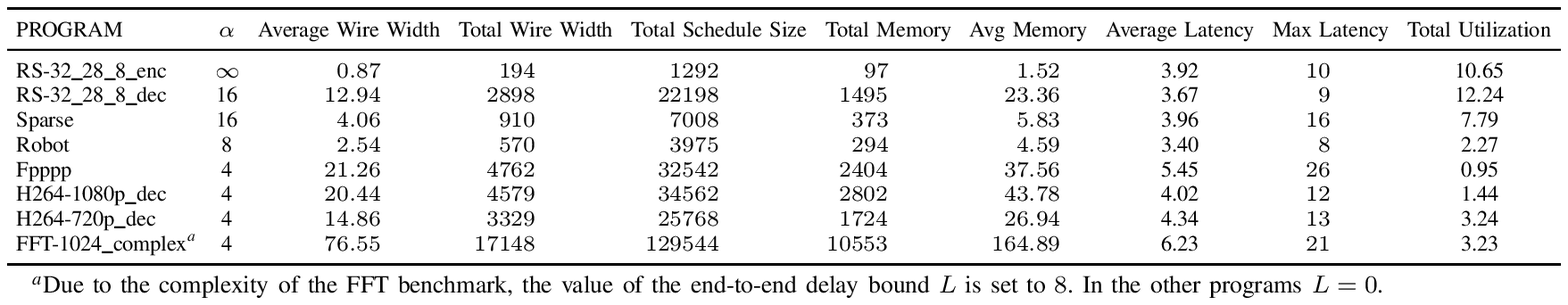}}
  \caption{Results with MCSL recorded traffic over an $8\times 8$ mesh with schedules
    of $8$ time slots per period}
\label{tbl:mesh 8x8}
\end{table}

\paragraph{Low Utilization with the MCSL Recorded Traffic.}  The ratio between the
utilized time slots and the total time slots is very small in the simulations
listed in Table~\ref{tbl:mesh 8x8}. The algorithm computes a static routing, and
therefore, to satisfy the real-time constraints must allocate bandwidth according to
worst case traffic. On the other hand, the benchmark is very bursty.  Indeed, in this
setting, the utilization is upper bounded as follows:
\[
\text{utilization}\leq \frac{\sum_{i,j} \text{average}(\bw(PE_i,PE_j))}
{\sum_{i,j} {\max}(\bw(PE_i,PE_j))}
\]
In Table~\ref{tbl:mcsl utilization}, we compare the utilization obtained during the simulations
of the NoC generated by the algorithm and the upper bounds on the utilization.
This comparison shows that the low utilization can not be avoided by static
schedulers that satisfy the real-time constraints of the MCSL recorded benchmarks.

\begin{table}
\centering
\begin{tabular}{c}
\includegraphics{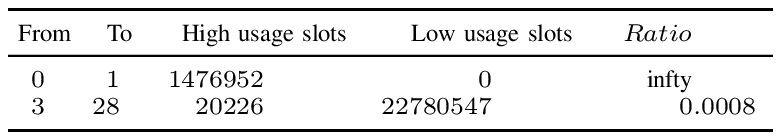}
\end{tabular}
  \caption{Results with MCSL recorded traffic over an $8\times 8$ mesh with schedules
    of $8$ time slots per period. }
  \label{tbl:mcsl utilization}
\end{table}
\section{Further Work}\label{sec:further}
Static routing does not adapt to changes in traffic. Three possible directions for
resolving this problem are: (1)~Introduce a network manager entity that gathers
information about the over- and under-utilizations of allocated bandwidth. New
routings and schedules can be computed from time to time by the network manager based
on the gathered information. The new schedules can be used to reconfigure the
switches and the NIs. Note that in such a setting, NoC link widths are fixed, although
one can re-orient wires of interconnections. (2)~Overlay dynamic routing over the
static routing. The idea is to mimic the way people use static schedules of trains
for their dynamic needs. (3)~Increase interconnection widths so that there are plenty
of time slots that are not allocated by the algorithm. These vacant time slots can be
used for a second dynamic routing protocol over the same network. Such a hybrid
solution can encapsulate the static and dynamic networks so that they do not mix although they use
the same infrastructure.


\bibliographystyle{plain}
\bibliography{IEEEabrv,refs}

\end{document}